\documentclass[12pt]{article}

\usepackage{newtxtext,newtxmath,color}

\usepackage{graphicx}

\usepackage[letterpaper,margin=1in]{geometry}

\renewenvironment{abstract}
	{\quotation}
	{\endquotation}

\date{}

\makeatletter
\renewcommand{\fnum@figure}{\textbf{Figure \thefigure}}
\renewcommand{\fnum@table}{\textbf{Table \thetable}}
\makeatother

\usepackage{scicite}

\usepackage{url}

\def\scititle{
Back Reaction of the Untwisting Solar Corona Scars Sunspots
}
\title{\bfseries \boldmath \scititle}

\author{
	Chen~Xing$^{1\ast}$,
	Xin~Cheng$^{1\ast}$,
	Guillaume~Aulanier$^{2,3}$,
	Mingde~Ding$^{1}$\and
	\small$^{1}$School of Astronomy and Space Science, Nanjing University, Nanjing, China.\and
	\small$^{2}$Sorbonne Université, Observatoire de Paris-PSL, École Polytechnique, IP Paris, CNRS,\and
	\small Laboratoire de Physique des Plasmas, Paris, France.\and
	\small$^{3}$Rosseland Centre for Solar Physics (RoCS), Institute of Theoretical Astrophysics, Universitetet i Olso, Oslo, Norway.\and
	\small$^\ast$Corresponding author. Email: chenxing@nju.edu.cn\and
	\small$^\ast$Corresponding author. Email: xincheng@nju.edu.cn
}

\begin{document} 

\maketitle

\begin{abstract} \bfseries \boldmath
The evolution of magnetic fields in the tenuous solar corona is predominantly governed by the motions of the underlying dense photosphere. Despite, coronal magnetic restructuring driven by magnetic reconnection between interacting coronal fields can sometimes react backwards to change photospheric magnetic fields. However, the mechanism of reactions remains undetermined. Here, we report the discovery of a back-reaction phenomenon: the untwisting of coronal loops that become twisted during reconnection in an eruption results in enhanced currents at the boundary of their footpoint away from the eruption, manifesting as the growth of a sunspot scar. It is revealed to arise from the Alfvénic reverse transfer of magnetic twist from the corona to the lower atmosphere, thanks to joint space observations and a magnetohydrodynamics simulation. These findings provide a viable and quantitative interpretation for the majority of puzzling photospheric changes associated with coronal mass ejections and/or flares and warn for unexpected magnetic field evolutions in sunspots and starspots.
\end{abstract}

\subsection*{Introduction}
\noindent
Magnetic fields are an essential component of the solar and stellar atmospheres. They are of paramount importance to power coronal heating, wind acceleration, and also explosions such as coronal mass ejections (CMEs) and flares \cite{Wiegelmann2014,Vidotto2018} that often give rise to solar energetic particles, geomagnetic and ionospheric storms, and Aurora Borealis \cite{Elovaara2007,Holman2011,Zharkova2011}. Given the low/high plasma beta ($\beta$) in the solar corona/photosphere \cite{Gary2001}, there is a consensus that coronal magnetic fields are line-tied to the photosphere and predominantly governed by photospheric motions \cite{Green2009,Liu2014}. Coronal magnetic fields are initially generated as flux tubes emerge through the photosphere due to magnetic buoyancy \cite{Magara2003,Liu2014}. As the emergence proceeds, an imbalance arises between highly twisted convection-zone fields and less twisted coronal fields. In consequence, shearing flows parallel to polarity inversion lines (PILs) \cite{Shimizu2014} and rotational motions of sunspots \cite{Yan2015,Vemareddy2016} are formed and twist up the emerged coronal fields into sheared arcades \cite{Fan2009,Liu2014}. Driven by either photospheric converging flows \cite{Green2009,Aulanier2010} or supergranular flows \cite{Antiochos2013,Liu2022}, they can be further twisted up via magnetic reconnection to form flux ropes lying above PILs \cite{Patsourakos2020}.

The non-potential fields in the vicinity of PILs, i.e., sheared arcades and flux ropes, carry large amounts of magnetic free energy. Due to the eruption and/or the interaction of the energized fields \cite{Somov1998,Aulanier2021}, magnetic reconnection rapidly converts stored free energy into kinetic and thermal energies and particle acceleration and simultaneously alters the connectivity of coronal magnetic fields (i.e., coronal magnetic restructuring) \cite{Antiochos1998,Aulanier2019}, producing CMEs and/or flares. Interestingly, for some events, the photospheric magnetic field near the PILs of source active regions becomes more horizontal \cite{Wang2002,Zharkova2005,Petrie2012} while that at the periphery of source regions turns more vertical \cite{Sudol2005,Wang2012,Sun2017}. It is even found that sunspots exhibit unconventional rotations as being swept by flare ribbons \cite{Bi2016,Liu2016}. Based on these observations, it is suggested that a change in photospheric magnetic fields could occur because of the back reaction (coined in \cite{Hudson2008}) of the coronal magnetic restructuring during CMEs/flares on the photosphere \cite{Wang2012,Liu2016}. However, there has been long-standing controversy regarding the mechanism of the back reaction, considering that it would be difficult for coronal magnetic fields, controlled by photospheric motions in common sense \cite{Green2009,Liu2014}, to reversely affect magnetic fields in the high-beta photosphere \cite{Gary2001}.

Here, we investigate a new back-reaction phenomenon that occurs after and remotely from a failed eruption. Different from previously reported back-reaction phenomena occurring during and at the site of CMEs/flares, it occurs on larger spatiotemporal scales. This enables us to resolve its unfolding and decipher its mechanism with a great level of details. It is found that the large-scale coronal loops become more twisted during the eruption but experience untwisting afterwards. The latter leads to the long-lasting growth of a scar in remote sunspot umbrae where the coronal loops are rooted, manifested as an enhancement of the current at the boundary of the loops' footpoint. Utilizing a three-dimensional (3D) magnetohydrodynamics (MHD) simulation, we not only reproduce the main observational characteristics but also reveal that the untwisting coronal fields grow the scar by a reverse transfer of magnetic twist from the corona to the lower atmosphere, in the form of Alfvén waves. These findings present a viable and quantitative interpretation for various photospheric changes induced by solar CME and flare events and warn for unexpected evolutions of magnetic fields in sunspots and starspots.

\subsection*{Results}
\subsubsection*{Event Overview}
\begin{figure}
\centering
\includegraphics[width=0.95\hsize]{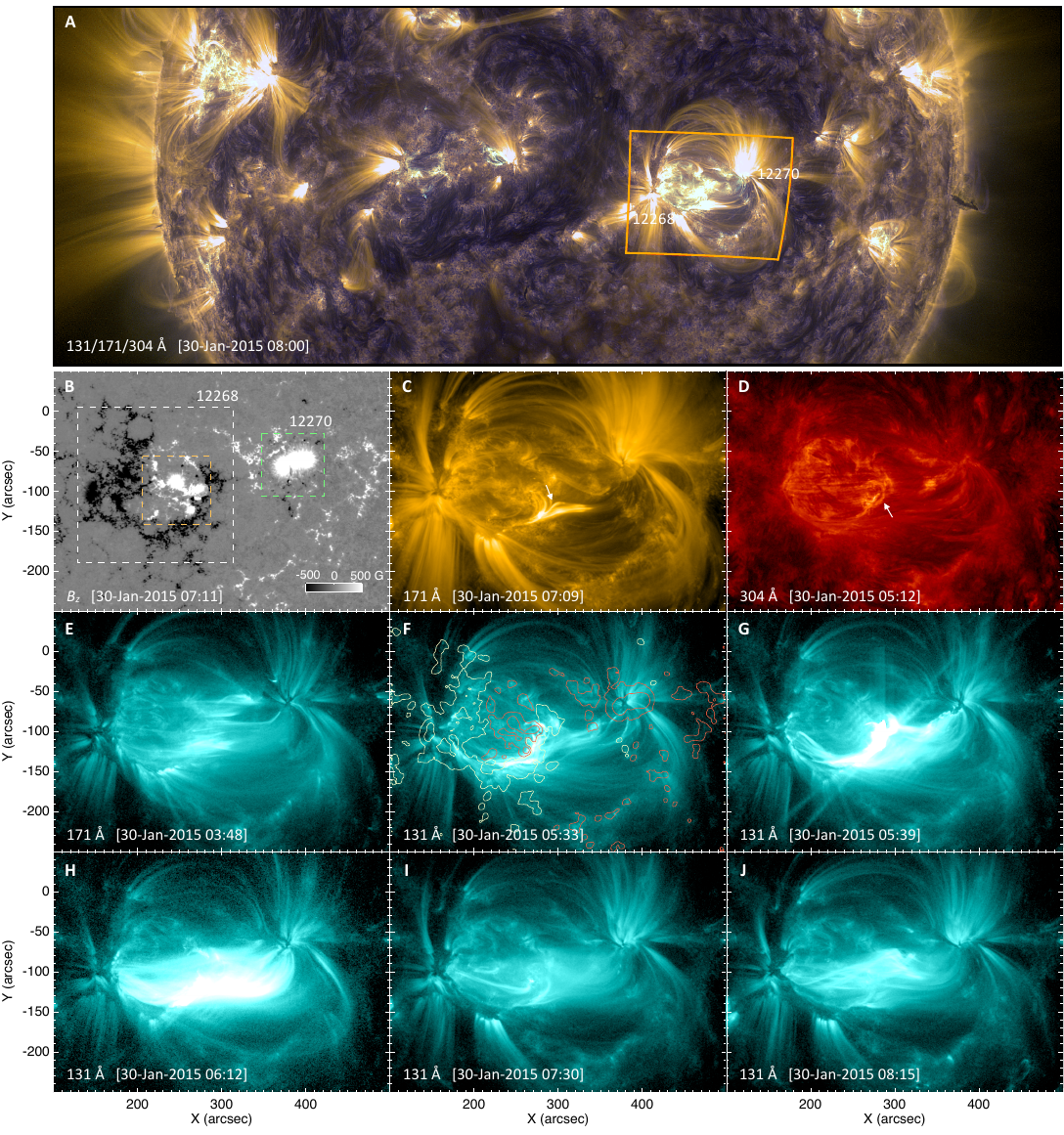}
\caption{\textbf{Overview of the failed eruption.} (\textbf{A}) Composite image of AIA 131 {\AA}, 171 {\AA}, and 304 {\AA}, showing the ARs 12268/70. The box marks the field of view (FOV) of panels B-J. (\textbf{B}) The distribution of $B_z$ in ARs 12268/70. The white, green, and orange boxes outline the AR 12268, AR 12270, and the positive polarity of AR 12268, respectively. (\textbf{C}) The AIA 171 \AA\ image of ARs 12268/70 with an arrow pointing to a bunch of loops overlying the ARs. (\textbf{D}) The AIA 304 \AA\ image before the eruption with an arrow pointing to the pre-eruptive filament. (\textbf{E})-(\textbf{J}) The AIA 131 \AA\ images before, during, and after the eruption. The red and yellow contours in panel F outline the positive and negative polarities, respectively.}
\label{fig1}
\end{figure}

On 30 January 2015, a major eruption occurred in the active region (AR) 12268 and spread to the AR 12270, at the west of the disk centre (Fig. \ref{fig1}A). As indicated by the Geostationary Operational Environmental Satellite (GOES) soft X-ray flux, an M1.7-class flare that accompanied the eruption started at 05:29 UT, peaked at 05:36 UT, and ended at 06:35 UT. The eruption was failed, considering no prominent coronal dimmings in 211 \AA\ around ARs 12268/70 and no CMEs appearing in the white-light coronagraph after the eruption (fig. S1). The magnetic topology of these two ARs is a fan-spine configuration \cite{Lau1990,Masson2009}, with its inner/outer spines anchored in the positive polarities of ARs 12268/70 and its fan separatrix surface rooted in the negative polarity of AR 12268 (Fig. \ref{fig1}B). The fan-spine topology is also illuminated by a bunch of coronal loops overlying two ARs, whose middle portion is cusp-shaped (Fig. \ref{fig1}C) and similar to the shape of field lines near the null point of the fan-spine structure \cite{Yang2020}.

The failed eruption originated from a pre-eruptive filament in AR 12268 (Fig. \ref{fig1}D). As the flare started, the erupting structure first appeared as a hot channel \cite{Cheng2011,Zhang2012} in 131 \AA, with its northern/southern footpoints anchored at the positive/negative polarities of AR 12268 (Fig. \ref{fig1}F). The hot channel soon extended west to AR 12270 and east to the easternmost part of AR 12268 (Fig. \ref{fig1}G). It later evolved into a diffuse structure connecting ARs 12268/70 (Fig. \ref{fig1}H) and then disappeared at 07:30 UT (Fig. \ref{fig1}I). Besides, the failed eruption produced not only a pair of flare ribbons in AR 12268, but also a remote hooked flare ribbon in AR 12270 (Fig. \ref{fig2}F,I). These observations indicate a magnetic reconnection between the erupting structure and the large-scale overlying field of ARs 12268/70, which caused changes in the magnetic connectivity and heating of the latter.

In addition to the brightening during the eruption, the overlying loops of ARs 12268/70 were also brightened twice in 131 \AA\ before and after the eruption, respectively. The first brightening occurred during 03:21-04:48 UT, initially appearing as a bright block around (300,-100) arcsec and then extending east and west into bright threads jointing two ARs (Fig. \ref{fig1}E). The second brightening occurred during 08:00-10:36 UT when a bright block first appeared around (300,-90) arcsec and then also evolved into bright overlying threads (Fig. \ref{fig1}J). As shown by the configurations of bright threads, the overlying field of ARs 12268/70 is less twisted before the eruption (Fig. \ref{fig1}E) but highly twisted after the eruption (Fig. \ref{fig1}J). This suggests that the eruptive structure injected magnetic twist into the overlying loops through reconnection during the eruption, thus transforming the overlying field from a weakly twisted structure to a highly twisted flux rope.

More interestingly, the coronal portion of twisted overlying loops showed an obvious untwisting after the eruption, which is particularly apparent in 131 \AA\ during 08:00-09:00 UT (Fig. \ref{fig2}A-E,J-L). Specifically, at 08:15 UT, the coronal portions of overlying loops were highly intertwined with each other, showing the feature of a twisted flux rope, with the twist number of some threads being about 1 turn (Fig. \ref{fig2}C). However, they clearly became less twisted at 08:42 UT (Fig. \ref{fig2}E). The untwisting phenomenon continued during 09:00-10:12 UT, while it was visible in 304 \AA\ rather than 131 \AA\ then (Fig. \ref{fig2}J-L).

\subsubsection*{Eruption-induced Growth of Sunspot Scar}
During and after the eruption, an arc-shaped structure appeared in the positive-polarity sunspot umbra of AR 12270, more than 70 Mm away from the origin of the failed eruption in AR 12268 (Fig. \ref{fig1} and Fig. \ref{fig3}). The length of this pronounced structure is comparable in magnitude to the length of the sunspot umbra along the $y$-direction (Fig. \ref{fig3}). Compared to its surrounding umbrae, the arc-shaped structure had a weaker vertical magnetic field ($B_z$; Fig. \ref{fig3}A-C) and a larger inclination angle of the magnetic field to the vertical direction (Fig. \ref{fig3}G-I). In addition, it had a negative/positive vertical current density ($J_z$) in its eastern/western part (Fig. \ref{fig3}J-L).

The joint observations show that the western footpoint of twisted overlying loops was located on the east of the arc-shaped structure after the eruption (Fig. \ref{fig2}B,D,G,H). In addition, the remote hooked flare ribbon was also located on the east of the arc-shaped structure (Fig. \ref{fig2}F,I), the former of which marked the footpoints of field lines formed through the reconnection between the erupting structure and overlying fields. Therefore, the arc-shaped structure is identified to be located at the boundary of the western footpoint of large-scale twisted overlying fields.

The properties of the arc-shaped structure in AR 12270 are similar to those of the sunspot scar \cite{Xing2024a} which is an arc-shaped structure with an inclined magnetic field in the sunspot umbra. The sunspot scar exhibits photospheric features similar to those of classical light bridges \cite{Solanki2003,Xing2024a}, while they possess distinctly different magnetic configurations. Classical light bridges are believed to comprise low-lying loops beneath canopy-shaped overlying fields \cite{Toriumi2015}. By contrast, the sunspot scar is located at the footpoint boundary of the flux rope. Its current on the side adjacent to the flux rope belongs to the direct current of the flux rope, while its opposite-sign current on the other side constitutes part of the return current \cite{Xing2024a}. In consequence, the arc-shaped structure observed here is recognized as a sunspot scar, with its negative current representing a part of the direct current of twisted overlying fields.

The sunspot scar in AR 12270 was substantially built up after the eruption. For example, compared to the sunspot scars at 07:10 UT and 09:34 UT, the sunspot scar at 11:34 UT was more pronounced in the $B_z$ map (Fig. \ref{fig3}A-C), exhibited an enhancement of the horizontal magnetic field ($B_h$; Fig. \ref{fig3}D-F), and had a stronger negative $J_z$ (Fig. \ref{fig3}J-L). The growth of the sunspot scar is further quantified by the evolution of the integral negative current in the scar (Fig. \ref{fig3}M and materials and methods S2). It is shown that the integral negative current first experienced a decrease by $\sim22\%$ from $\sim$05:34 UT to $\sim$07:10 UT (Fig. \ref{fig3}M). Later, it increased by $\sim73\%$ from $\sim$07:34 UT to $\sim$11:34 UT (Fig. \ref{fig3}M), which clearly represents the long-lasting growth of the sunspot scar after the eruption.

Importantly, the evolution of the sunspot scar is highly relevant to the coronal dynamics in ARs 12268/70 during and after the eruption. Firstly, the integral negative current in the sunspot scar decreased since the onset of the eruption and until about half an hour after the end of the flare (Fig. \ref{fig1} and Fig. \ref{fig3}M). Secondly, the integral negative current of the sunspot scar increased when the coronal portion of large-scale twisted overlying loops anchored close to the scar exhibited an untwisting (Fig. \ref{fig2}, Fig. \ref{fig3}M, and Supplementary Text).

Such a close spatiotemporal connection between the coronal dynamics and the evolution of the sunspot scar indicates the back reaction of the former on the latter and could be explained in the following scenario. During the eruption, as the erupting twisted structure reconnects with the overlying loops of ARs 12268/70, the magnetic twist is transferred from the former to the latter, making the overlying loops, to be specific, their coronal portion, more twisted (Fig. \ref{fig1}). At the same time, the erupting structure also stretches the overlying loops, not only their coronal portion but also their feet in the lower atmosphere. This leads to a decrease in the twist number per unit length and thus a reduction of the direct current at the feet of loops, as represented by the decrease of the integral negative current (direct current) in the sunspot scar (Fig. \ref{fig3}M).

After the eruption, the large-scale twisted overlying loops gradually relax, with magnetic twist being redistributed along loops. Naturally, the highly twisted coronal portion of loops becomes less twisted, appearing as the untwisting of coronal loops in observations (Fig. \ref{fig2}). Meanwhile, the less twisted lower-atmosphere portion of loops becomes more twisted, resulting in an increase of the direct current at their western footpoint in the remote sunspot. Such an increase is particularly pronounced at the footpoint boundary \cite{Janvier2014}, which is manifested as the enhancement of the integral negative current in the sunspot scar, i.e., the growth of the scar (Fig. \ref{fig3}). In other words, the untwisting of twisted coronal loops reacts backwards to grow the sunspot scar at their footpoint boundary via a reverse transfer of magnetic twist from the corona along loops to the lower atmosphere.

\subsubsection*{MHD Simulation}
To confirm the scenario in observations, we performed an observationally inspired MHD simulation of a failed eruption. The initial magnetic field, mimicking the observed pre-eruptive field, is composed of a potential field in the fan-spine configuration and a modified Titov-Démoulin (TD) flux rope \cite{Titov1999} embedded beneath the fan separatrix surface (Fig. \ref{fig4}A and materials and methods S3). The unequilibrium flux rope immediately erupts at the beginning and then reconnects with the overlying fan-spine field, giving rise to a failed eruption (fig. S2). Because of reconnection, one footpoint of the flux rope progressively jumps from the positive polarity below the fan separatrix surface to the remote positive polarity, close to the footpoint of the outer spine (Fig. \ref{fig4} and fig. S2). Meanwhile, the other footpoint of the flux rope drifts to the region where the footprints of the original fan separatrix surface were (Fig. \ref{fig4} and fig. S2).

After the eruption, a sunspot scar, without being pre-set in the simulation, naturally appears at the upper left boundary of the flux rope footpoint in the remote positive polarity (Fig. \ref{fig5}D-F). Compared to either the situation at the same location before the eruption or the surrounding sunspot umbrae, the modeled sunspot scar shows a smaller $B_z$ (Fig. \ref{fig5}A,D), a larger $B_h$ (Fig. \ref{fig5}B,E), and a larger inclination angle of magnetic field (Fig. \ref{fig5}C,F), highly resembling observations. There is a strong negative $J_z$ in the sunspot scar at the side close to the flux rope and a strong positive $J_z$ at the other side (Fig. \ref{fig5}G). The increase of the integral negative current in the scar, as represented by those of currents in three boxes in the negative-current region of the scar (fig. S3), is similar to that in observations (Fig. \ref{fig3}M) and shows the growth of the modeled sunspot scar. The twisting of magnetic fields, quantified by a parameter $\tau=\boldsymbol{J}\cdot\boldsymbol{B}/B^2$ (Fig. \ref{fig5}H), exhibits a distribution similar to that of $J_z$, indicating highly twisted fields in the sunspot scar (Fig. \ref{fig5}G,H).

We study the formation of the modeled sunspot scar by the evolution of the magnetic twist along a field line ($L_1$) anchored inside the sunspot scar (Fig. \ref{fig4}B). The field line $L_1$ is located below the outer spine before the eruption (Fig. \ref{fig4}A). In a short period during the eruption, two groups of magnetic twist are successively generated along $L_1$, one induced by the disturbance of the eruption and the other injected from the erupting flux rope via magnetic reconnection (fig. S4A,D and materials and methods S4). Then, in a long period after the eruption, the twist is transferred along $L_1$ in the form of Alfvén waves (fig. S4F,G), reversely from its coronal portion to its feet (fig. S4D). Upon arriving at the low atmosphere, the magnetic twist affects the local distribution of current density and magnetic field, leading to the formation of the sunspot scar (Fig. \ref{fig5}). By further investigating the evolution of a field line ($L_2$) rooted outside the sunspot scar (Fig. \ref{fig4}B and fig. S4B), it is determined that the reconnection-induced magnetic twist, rather than the disturbance-induced one, plays a key role in forming and maintaining the sunspot scar (fig. S4D,E). These results fully confirm the scenario in observations, i.e., that the untwisting of the reconnected coronal magnetic field formed by the failed eruption induces a reverse transfer of magnetic twist downwards from the corona in the form of Alfvén waves, which reacts on the lower atmosphere and forms the sunspot scar.

\subsection*{Discussion}
It is generally believed that the photospheric motions control the evolution of coronal magnetic fields to a large extent. An opposite process, i.e., the back reaction of the coronal magnetic restructuring on the photosphere, although indicated to exist, is still unclear in terms of its mechanism. Here, we report a phenomenon that the untwisting of large-scale coronal magnetic fields leads to a long-lasting growth of a scar in sunspot umbrae. It provides another piece of evidence for the existence of back reaction in the solar atmosphere. Moreover, the new back-reaction phenomenon occurs after and away from the eruption, rather than during and at the site of CMEs/flares as reported previously \cite{Petrie2012,Wang2012,Liu2016}, indicating a more common presence of back reactions. This difference may also explain why the current evolution during the back reaction here is so gradual that it deviates from the step-like evolution in previous back reactions during CMEs/flares \cite{Petrie2012,Sun2017}, as it is driven by a post-eruption long-lasting untwisting process spanning two ARs rather than rapid CME/flare processes limited within their source regions.

More importantly, we determine that the untwisting coronal fields react backwards on the sunspot through an Alfvénic reverse transfer of magnetic twist. Other mechanisms during flares, such as the thermal pressure impulse induced by chromospheric evaporation or radiative back-warming \cite{Fisher2012} and the accelerated particle capable of causing variations in electric and magnetic fields \cite{Munoz2018,Khabarova2020,Zharkova2021}, might also contribute to the back reaction. However, they most likely do not play a major role in the event studied here because the scar grows in a long period after the eruption. The mechanism, i.e., reverse transfer of magnetic twist from the corona to the photosphere, is disclosed in a failed eruption event, where considerable twist was transferred to the corona through magnetic reconnection to create a twist imbalance between the corona and photosphere that established preconditions for the reverse twist transfer. It is thus expected that such a mechanism is also able to occur in successful eruptions or even flares without eruptions, provided a similar twist imbalance generated via reconnection in these events. In different events, the reverse twist transfer may give rise to distinct photospheric changes, depending upon many factors including the magnetic configuration and spatiotemporal scale at which the reverse twist transfer occurs. Specifically, in addition to the growth of sunspot scars at the boundary of flux rope footpoints, this mechanism may also lead to the enhancement of photospheric currents/horizontal magnetic fields near PILs during eruptions \cite{Petrie2012,Sun2017}, as the twist reversely propagates from the cusp-shaped top of reconnection-formed flare loops to the loops' feet. It may even explain the unconventional rotation of sunspots during flares \cite{Bi2016,Liu2016}, considering the potential twist imbalance around the photosphere caused by the reverse twist transfer from the corona to photosphere. Such a mechanism differs from the reconnection-driven contraction of flare loops \cite{Barczynski2019} and the loop contraction due to a momentum conservation of eruptions \cite{Fisher2012}, the latter two of which apply only to the enhanced horizontal field near PILs \cite{Petrie2012,Sun2017} as revealed in a CME simulation \cite{Barczynski2019} and suggested hypothetically \cite{Fisher2012}, respectively. In addition, our result supports the prediction that the back reaction proceeds as the Alfvén waves launched from non-force-free coronal fields propagate downwards to change photospheric magnetic fields \cite{Aulanier2016}, which is difficult to be confirmed in many back-reaction events whose durations are too short to determine Alfvén waves \cite{Wang2002,Castellanos2018}.

The discovered reverse transfer of magnetic twist, combined with the forward propagation of twist from the photosphere to the corona \cite{Pariat2009,Torok2009,Wyper2016}, establishes a framework for the magnetic twist circulation in the solar atmosphere. Within the framework, the magnetic twist injected into the corona by eruptive structures could propagate reversely to the surface in source regions or other ARs, producing photospheric responses such as scars in sunspot umbrae \cite{Xing2024a} or enhanced horizontal fields at inner penumbrae near PILs \cite{Sun2017}. Besides, for most of the time in the absence of eruptions, the forward and reverse transfer of the twist along coronal loops could exchange magnetic helicity between distinct ARs \cite{Yang2009}, changing the distribution of photospheric magnetic fields in sunspots on a large scale. These changes in photospheric fields could be even more substantial in other active stars, considering the more drastic CMEs, flares, and evolutions of ARs in these stars compared to the Sun's \cite{Drake2000,Namekata2022}. Therefore, it is warned that, due to the additional contribution of the magnetic twist circulation in atmospheres, the magnetic fields in sunspots and starspots will exhibit an evolution other than that expected as governed by the convection and subsurface flux emergence alone.


\subsection*{Materials and Methods}
\subsubsection*{S1. Observation Data}
We mainly use extreme ultraviolet (EUV)/UV images from Atmospheric Imaging Assembly (AIA; \cite{Lemen2012}), and Spaceweather HMI Active Region Patch (SHARP) cylindrical equal-area (CEA) coordinate vector magnetic field maps \cite{Bobra2014} and helioprojective-Cartesian coordinate vector magnetic field maps from Helioseismic and Magnetic Imager (HMI; \cite{Scherrer2012}), both of which are on board Solar Dynamics Observatory (SDO; \cite{Pesnell2012}). The cadence of EUV/UV images is 12/24 seconds, and the cadence of magnetic field maps is 12 minutes. We also take advantage of white-light images from Large Angle Spectroscopic Coronagraph (LASCO; \cite{Brueckner1995}) on board Solar and Heliospheric Observatory (SOHO; \cite{Domingo1995}).

\subsubsection*{S2. Measurement of Current of Sunspot Scar}
We measure the integral negative current of the sunspot scar in AR 12270, every 12 minutes during 04:22-14:22 UT on 30 January 2015, following the pipeline below.
\begin{itemize}
\item[1.] First, we derive the distribution of the vertical current density ($J_z$) in AR 12270 using the SHARP CEA-coordinate vector magnetic field. Considering the uncertainty in magnetic field measurement, we only calculate $J_z$ in regions with both horizontal and vertical field strength larger than 100 G.
\item[2.] The integral negative current is derived by integrating $J_z$ in the negative-current region of the sunspot scar, which is identified with a negative $J_z$ stronger than a fixed threshold. To determine the threshold, we first calculate the average ($J_a^i<0$) and the standard deviation ($J_{sd}^i>0$) of negative $J_z$ in the region ($51.02\le x\le56$, $-11\le y\le-7.88$, unit is CEA degree; a little larger than the FOV of Fig. \ref{fig3}A-L) for each $J_z$ map (labeled as $i=1,2,3,...51$) during 04:22-14:22 UT. The threshold ($J_c$) is then given by:
\begin{equation}
J_c=\frac{\sum_{i=1}^{51}(J_a^i-J_{sd}^i)}{51}.
\end{equation}
Using this threshold, $J_c\approx-0.011\ \textup{A}/\textup{m}^2$, we can identify regions where the negative $J_z$ is considerably stronger than the average of negative $J_z$ in the positive polarity of AR 12270.
\item[3.] Then, we use this threshold to determine the negative-current region in the sunspot scar every 12 minutes. In most cases, we identify only one region and then calculate the integral of the negative current within. However, in a few cases, we also identify some negative-current patches adjacent but disjoint to the major negative-current region. In this case, we calculate the integral twice, once including and once excluding these patches; the integral negative current shown in Fig. \ref{fig3}M is then taken as an average of two measurements.
\end{itemize}

The error of the integral negative current $\sigma_I$, marked by the error bar in Fig. \ref{fig3}M, arises from two sources. The first component is the standard deviation between two measurements, denoted as $\sigma_{I_{sd}}$. For cases measured only once, $\sigma_{I_{sd}}$ is set to zero. The second component originates from the measurement error in the horizontal magnetic field, noted as $\sigma_{I_B}$. As these two components are independent, the total error is given by:
\begin{equation}
\sigma_I = \sqrt{\sigma_{I_{sd}}^2+\sigma_{I_B}^2}.
\end{equation}
The parameter $\sigma_{I_B}$ is calculated as follows:
\begin{itemize}
\item[1.] At each pixel of determined negative-current region in sunspot scar, the error of the $x$-component of magnetic field $\sigma_{B_x}$ and that of the $y$-component of magnetic field $\sigma_{B_y}$ are provided by the SHARP data. The error of the current in each pixel arising from the error of horizontal field $\sigma_{I_i}$ is then given by:
\begin{equation}
\sigma_{I_i} = \frac{1}{\mu_0}\sqrt{(\Delta x\sigma_{B_x})^2+(\Delta y\sigma_{B_y})^2},
\end{equation}
where the subscript $i$ marks the sequence number of the pixel in the negative-current region. The parameters $\Delta x$ and $\Delta y$ represent the size of the pixel in $x$ and $y$ directions, respectively. $\mu_0$ is vacuum permeability.
\item[2.] For cases in which the negative-current region is only determined once, the error of the integral negative current in the sunspot scar due to the horizontal field error, i.e., $\sigma_{I_B}$, is given by:
\begin{equation}
\sigma_{I_B} = \sqrt{\sum_{i=1}^{n}\sigma_{I_i}^2},
\end{equation}
where $n$ refers to the number of pixels in the negative-current region. For other cases where the negative-current region is determined twice, this error is given by:
\begin{equation}
\sigma_{I_B} = \frac{1}{2}\sqrt{\sum_{i=1}^{n_1}\sigma_{I_i}^2+\sum_{i=1}^{n_2}\sigma_{I_i}^2},
\end{equation}
where $n_1$ and $n_2$ denote the numbers of pixels in the negative-current regions determined twice, respectively.
\end{itemize}

\subsubsection*{S3. Simulation Setup}
We performed a 3D observationally inspired zero-beta MHD simulation of a failed eruption to mimic the observations with the code MPI-AMRVAC \cite{Xia2018}. The simulation solves the following equations in Cartesian coordinates:
\begin{equation}\label{eq1}
\frac{\partial\rho}{\partial t}+\nabla\cdot(\rho\boldsymbol{v}) = 0
\end{equation}
\begin{equation}\label{eq2}
\begin{aligned}
\frac{\partial(\rho\boldsymbol{v})}{\partial t}+\nabla\cdot(\rho\boldsymbol{v}\boldsymbol{v}+\frac{\boldsymbol{B}^2}{2\mu_0}\boldsymbol{I}-\frac{\boldsymbol{B}\boldsymbol{B}}{\mu_0}) = 0
\end{aligned}
\end{equation}
\begin{equation}\label{eq3}
\frac{\partial \boldsymbol{B}}{\partial t}+\nabla\cdot(\boldsymbol{v}\boldsymbol{B}-\boldsymbol{B}\boldsymbol{v}+\psi\boldsymbol{I}) = 0
\end{equation}
\begin{equation}\label{eq4}
\nabla\times\boldsymbol{B} = \mu_0\boldsymbol{J}
\end{equation}
\begin{equation}\label{eq5}
\frac{\partial\psi}{\partial t}+c_h^2\nabla\cdot\boldsymbol{B} = -\frac{c_h^2}{c_p^2}\psi.
\end{equation}
Here, $\rho$, $\boldsymbol{v}$, $\boldsymbol{B}$, $\boldsymbol{J}$ represent the mass density, velocity, magnetic field, and current density, respectively. $\psi$ is the generalized Lagrange multiplier (GLM) in the GLM method \cite{Dedner2002} which maintains the $\nabla\cdot\boldsymbol{B}=0$ condition, and $c_h$ and $c_p$ are constants. The equations are solved dimensionlessly, and the units to dimensionalize the length, time, mass density, velocity, and magnetic field strength are 10 Mm, 85.87 s, $2.34\times10^{-15}$ g cm$^{-3}$, 116.45 km s$^{-1}$, and 2 G, respectively (in this paper, parameters are dimensionless unless indicated). We do not set an explicit resistivity, but the numerical resistivity effectively facilitates the magnetic reconnection.

The simulation domain is a cube of $-7\le x\le11$, $-6\le y\le6$, and $0\le z\le14$. The domain is resolved by 192 symmetric-stretched grids with a stretched ratio of 1.021 in $x$ direction, 160 symmetric-stretched grids with a stretched ratio of 1.021 in $y$ direction, and 112 unidirectional-stretched grids with a stretched ratio of 1.0214 in $z$ direction. Benefiting from the stretched grids, the finest resolutions of the domain are about 300 km in all directions, comparable to that of \textit{SDO}/AIA. To achieve an accurate and stable simulation, we use the HLL scheme, the third-step Runge-Kutta time discretization method, the fifth-order weno5-limited reconstruction \cite{Jiang1996,Shu2009}, the GLM $\nabla\cdot\boldsymbol{B}$ cleaning method, and the magnetic field splitting method \cite{Tanaka1994,Xia2018}.

The initial distribution of mass density is derived from a stratified fully-ionized atmosphere including the chromosphere, transition region, and corona. The temperature of the stratified atmosphere is set to:
\begin{equation}\label{eq6}
T(z) = \left\{
\begin{aligned}
T_{ch}+\frac{1}{2}(T_{co}-T_{ch})(\tanh(\frac{z-h_{tr}-0.027}{w_{tr}})+1) & & & & & z\le h_{tr} \\
(\frac{7}{2}\frac{F_c}{\kappa}(z-h_{tr})+T_{tr}^{7/2})^{2/7} & & & & & z> h_{tr},
\end{aligned}
\right.
\end{equation}
where $T_{ch}=8\times10^{-3}$, $T_{tr}=0.16$, $T_{co}=1.5$, $h_{tr}=0.2$, $w_{tr}=0.02$, $F_c=0.054$, and $\kappa=0.22$. The gravity of the stratified atmosphere is set to:
\begin{equation}
g=g_0R_{sun}^2/(R_{sun}+z)^2,
\end{equation}
where $g_0=-0.20$ and $R_{sun}=69.55$. The mass density of the stratified atmosphere is then derived under the hydrostatic assumption and with $\rho_{z=0}=7.42\times10^{3}$ as input.

The initial magnetic field is composed of two parts. The first part is a potential field in the fan-spine configuration:
\begin{equation}\label{eq7}
\begin{gathered}
B_x(t=0)=\Sigma_{m=1}^4c_m(x-x_m)r_m^{-3} \\
B_y(t=0)=\Sigma_{m=1}^4c_m(y-y_m)r_m^{-3} \\
B_z(t=0)=\Sigma_{m=1}^4c_m(z-z_m)r_m^{-3} \\
r_m=\sqrt{(x-x_m)^2+(y-y_m)^2+(z-z_m)^2},
\end{gathered}
\end{equation}
where ($c_1=28.2$, $x_1=4.66$, $y_1=0$, $z_1=-2.2$), ($c_2=-23.3$, $x_2=0.89$, $y_2=0$, $z_2=-1.8$), ($c_3=41.2$, $x_3=-0.94$, $y_3=0$, $z_3=-1.1$), and ($c_4=-56.6$, $x_4=-1.24$, $y_4=0$, $z_4=-2.2$). The second part is a modified TD flux rope inserted underneath the fan separatrix surface, whose magnetic field is composed of $B_I$ and $B_\theta$ of the TD flux rope \cite{Titov1999}. In the absence of the $B_q$ component \cite{Titov1999}, the flux rope is initially in a non-equilibrium state, which facilitates its eruption. The parameters of the modified TD flux rope are fine-tuned to ensure that it can be ejected and reconnect with a great amount of overlying fields during the eruption, but ultimately the eruption fails, well mimicking the observations. The axis of the TD flux rope is set in the plane $x=0$, with its apex reaching $z=0.6$ and two footpoints anchored around $(x,y,z)=(0,\pm0.75,0.015)$. Other parameters of the TD flux rope are $R=0.8$, $a=0.3$, $d=0.2$, $I=-6.313$, and $I_0=0.675$. Contributed by the fan-spine overlying field and the modified TD flux rope, the magnetic field strength in the plane $z=0$ is up to about 40 G.

The initial velocity and parameter $\psi$ are set to zero in the whole physical domain.

The first layer of the physical domain at the bottom boundary is set to be line-tied. The velocity and $\psi$ at this layer are set to zero, and the vertical magnetic field $B_z$ is fixed at this layer correspondingly. In the bottom ghost cells, $\boldsymbol{v}$, $\boldsymbol{B}$, and $\psi$ are set to fulfill the line-tied condition by following the setups in \cite{Xing2024b}, while the mass density is fixed to the initial value. For the other five boundaries, the setups of $\boldsymbol{v}$, $\boldsymbol{B}$, and $\psi$ are the same as those in \cite{Xing2024b}, while the mass density is derived by the second-order zero-gradient extrapolation.

\subsubsection*{S4. Analyses of Modeled Sunspot Scar}
In order to mimic the condition of $\beta\ge1$ in the photosphere, many numerical simulations of solar eruptions adopt a line-tied bottom boundary condition by imposing velocities at the line-tied layer  \cite{Aulanier2010,Jiang2021}. In our simulation, the line-tied layer is forced to be no-flow, leading to no change in $B_z$ there. However, this also means that the sunspot scar is not able to be present in this layer. Therefore, we choose  a horizontal surface at $z=0.293$ that is a little above the line-tied layer to study the sunspot scar (Fig. \ref{fig4} and Fig. \ref{fig5}). On the one hand, this surface ($z=0.293$) is close enough to the bottom so that the magnetic field here is still subject to the underlying line-tied effect. On the other hand, the magnetic field on this surface can also respond to the overlaying coronal evolution. Consequently, the evolution of the magnetic field on such a surface is able to reflect the back reaction of coronal dynamics on the magnetic field in the dense lower atmosphere. In this work, we refer to the surface at $z=0.293$ as the ``observation layer''.

To figure out how the magnetic twist/current in the sunspot scar is formed, we study the evolution of two representative field lines anchored around the scar. The first field line, referred to as $L_1$, is located below the outer spine before the eruption (see the green field line in Fig. \ref{fig4}A). By continuously tracing $L_1$ from its fixed positive-polarity footpoint at the line-tied layer (fig. S4A), we find that $L_1$ evolves into a twisted flux rope field line after the eruption (Fig. \ref{fig4}B), with its positive-polarity footpoint located in the negative-current/twist region of the sunspot scar (Fig. \ref{fig5}G,H).

We analyze the morphological evolution of $L_1$ (fig. S4A) and also the evolution of magnetic twist along it (fig. S4D), the latter of which is quantified by the parameter $\tau=\boldsymbol{J}\cdot\boldsymbol{B}/B^2$. fig. S4D shows the distribution of $\tau$ along $L_1$, with its positive-polarity footpoint at the line-tied layer as the origin; the blue dashed line marks the intersection of $L_1$ and the observation layer. Before the eruption, there is no magnetic twist along $L_1$ (fig. S4D). As the eruption occurs, $L_1$ is first disturbed by the eruption at $t=2$, with its coronal portion deviating from its initial state while its negative-polarity footpoint remaining almost unchanged (fig. S4A). Specifically, the disturbance produces a pair of twisted structures with negative and positive magnetic twist (indicated by two arrows in fig. S4A), respectively, along $L_1$ at $t=2$ (fig. S4D). During $2\le t\le4$, this pair of magnetic twist is transferred along $L_1$ toward its positive-polarity footpoint (fig. S4A,D).

In addition, $L_1$ is found to be reconnected with the erupting flux rope during $2\le t\le4$, leading to an apparent drift of its negative-polarity footpoint (fig. S4A). The left portion of $L_1$ becomes more twisted at $t=4$ compared to that at $t=2$ (fig. S4A), as the flux rope injects a strong negative twist (denoted by yellow arrows in fig. S4B,D) into $L_1$ through the reconnection. Then, as $L_1$ relaxes, two groups of twist (i.e., disturbance-induced twist and reconnection-induced twist) are gradually transferred towards the positive-polarity footpoint of $L_1$ during $4\le t\le10$ (fig. S4A,D). A negative magnetic twist first arrives at the observation layer around $t=10$ and then stays at this layer in a long period, at least until $t=20$ (e.g., see that at $t=14$ in fig. S4D). Meanwhile, the negative-current/twist region of the sunspot scar correspondingly appears at the observation layer, around the footpoint of $L_1$ (e.g., Fig. \ref{fig5}G,H).

It should be noted that there are two structures with negative twist along $L_1$, one induced by the eruption disturbance while the other injected by the reconnection. To uncover which one results in the long-lasting negative twist in the sunspot scar, we trace the other field line (hereafter referred to as $L_2$) with the same method as we do for $L_1$. $L_2$ is lower than $L_1$ at $t=0$ (Fig. \ref{fig4}A), and it is anchored outside the sunspot scar at $t=14$ (Fig. \ref{fig4}B and Fig. \ref{fig5}G). Unlike $L_1$, $L_2$ does not experience the reconnection (fig. S4B). The magnetic twist is zero along $L_2$ at $t=0$ (fig. S4E). The eruption also disturbs $L_2$ and induces a pair of twisted structures with negative and positive twist, respectively, along $L_2$ at $t=2$ (fig. S4E). Then, this pair of twist is transferred along $L_2$ towards its positive-polarity footpoint during $2\le t\le4$, and it induces a negative twist at the observation layer at $t=4$ (fig. S4E). However, the magnetic twist at the observation layer soon changes to being positive at $t=8$, and it finally diminishes close to zero at a later time (e.g., $t=12$ and $t=14$; fig. S4E).

By comparing the evolutions of $L_1$ and $L_2$, it is concluded that the negative twist induced by the eruption disturbance is not able to maintain the negative twist in the sunspot scar at the observation layer. Therefore, the negative twist/current in the sunspot scar is formed and maintained as the negative twist injected into the overlying loops (e.g., $L_1$) through the reconnection is reversely transferred from the coronal portion of loops to their feet in the lower atmosphere.

Finally, we study how the reconnection-induced twist is transferred along the loops. During $4\le t\le4.1$, $L_1$ changes slightly and its negative-polarity footpoint drifts a little distance (fig. S4C), indicating that it does not experience a substantial reconnection process in this period. The evolution of $\tau$ along $L_1$ is shown in fig. S4F. The gray box marks the leading front of the negative twist injected by the reconnection (fig. S4F), which is also pointed out by the red arrow in fig. S4B. During $4\le t\le4.1$, the leading front is transferred towards the positive-polarity footpoint of $L_1$ (fig. S4F,G), in agreement with the propagation of entire reconnection-induced negative twist. We further derive a shifted distribution of $\tau$ along $L_1$ at $t=4.05$ (black dashed curve in fig. S4G) by shifting the distribution of $\tau$ at $t=4$ (black solid curve in fig. S4G) towards the positive-polarity direction with the local Alfvén wave speed for 0.05 time unit (the local Alfvén wave speed is derived by averaging the speeds at $t=4$ and $t=4.05$). Similarly, we also derive a shifted distribution of $\tau$ at $t=4.1$ (red dashed curve in fig. S4G) by shifting the distribution of $\tau$ at $t=4.05$ (red solid curve in fig. S4G). The black (red) dashed curve basically matches the red (blue) solid curve in the range of [97.5,101.5] ([98,100]), indicating that the reconnection-induced twist is transferred along the loops in the form of Alfvén waves (fig. S4G).


\clearpage 

\bibliography{back_reaction} 

\begin{thebibliography}{10}
\providecommand{\url}[1]{\texttt{#1}}
\expandafter\ifx\csname urlstyle\endcsname\relax
  \providecommand{\doi}[1]{doi:\discretionary{}{}{}#1}\else
  \providecommand{\doi}{doi:\discretionary{}{}{}\begingroup
  \urlstyle{rm}\Url}\fi

\bibitem{Wiegelmann2014}
T.~{Wiegelmann}, J.~K. {Thalmann}, S.~K. {Solanki}, {The magnetic field in the
  solar atmosphere}. \emph{\aapr} \textbf{22}, 78 (2014),
  \doi{10.1007/s00159-014-0078-7}.

\bibitem{Vidotto2018}
A.~A. {Vidotto}, {Stellar Coronal and Wind Models: Impact on Exoplanets}, in
  \emph{Handbook of Exoplanets}, H.~J. {Deeg}, J.~A. {Belmonte}, Eds., p.~26
  (2018), \doi{10.1007/978-3-319-55333-7_26}.

\bibitem{Elovaara2007}
J.~{Elovaara}, {Finnish Experiences with Grid Effects of GIC'S}, in \emph{Space
  Weather : Research Towards Applications in Europe 2nd European Space Weather
  Week (ESWW2)}, J.~{Lilensten}, Ed., vol. 344 of \emph{Astrophysics and Space
  Science Library} (2007), p. 311, \doi{10.1007/1-4020-5446-7_27}.

\bibitem{Holman2011}
G.~D. {Holman}, \emph{et~al.}, {Implications of X-ray Observations for Electron
  Acceleration and Propagation in Solar Flares}. \emph{\ssr}
  \textbf{159}~(1-4), 107--166 (2011), \doi{10.1007/s11214-010-9680-9}.

\bibitem{Zharkova2011}
V.~V. {Zharkova}, \emph{et~al.}, {Recent Advances in Understanding Particle
  Acceleration Processes in Solar Flares}. \emph{\ssr} \textbf{159}~(1-4),
  357--420 (2011), \doi{10.1007/s11214-011-9803-y}.

\bibitem{Gary2001}
G.~A. {Gary}, {Plasma Beta above a Solar Active Region: Rethinking the
  Paradigm}. \emph{\solphys} \textbf{203}~(1), 71--86 (2001),
  \doi{10.1023/A:1012722021820}.

\bibitem{Green2009}
L.~M. {Green}, B.~{Kliem}, {Flux Rope Formation Preceding Coronal Mass Ejection
  Onset}. \emph{\apjl} \textbf{700}~(2), L83--L87 (2009),
  \doi{10.1088/0004-637X/700/2/L83}.

\bibitem{Liu2014}
Y.~{Liu}, \emph{et~al.}, {Magnetic Helicity in Emerging Solar Active Regions}.
  \emph{\apj} \textbf{785}~(1), 13 (2014), \doi{10.1088/0004-637X/785/1/13}.

\bibitem{Magara2003}
T.~{Magara}, D.~W. {Longcope}, {Injection of Magnetic Energy and Magnetic
  Helicity into the Solar Atmosphere by an Emerging Magnetic Flux Tube}.
  \emph{\apj} \textbf{586}~(1), 630--649 (2003), \doi{10.1086/367611}.

\bibitem{Shimizu2014}
T.~{Shimizu}, B.~W. {Lites}, Y.~{Bamba}, {High-speed photospheric material flow
  observed at the polarity inversion line of a {\ensuremath{\delta}}-type
  sunspot producing an X5.4 flare on 2012 March 7}. \emph{\pasj} \textbf{66},
  S14 (2014), \doi{10.1093/pasj/psu089}.

\bibitem{Yan2015}
X.~L. {Yan}, \emph{et~al.}, {The Formation and Magnetic Structures of
  Active-region Filaments Observed by NVST, SDO, and Hinode}. \emph{\apjs}
  \textbf{219}~(2), 17 (2015), \doi{10.1088/0067-0049/219/2/17}.

\bibitem{Vemareddy2016}
P.~{Vemareddy}, X.~{Cheng}, B.~{Ravindra}, {Sunspot Rotation as a Driver of
  Major Solar Eruptions in the NOAA Active Region 12158}. \emph{\apj}
  \textbf{829}~(1), 24 (2016), \doi{10.3847/0004-637X/829/1/24}.

\bibitem{Fan2009}
Y.~{Fan}, {The Emergence of a Twisted Flux Tube into the Solar Atmosphere:
  Sunspot Rotations and the Formation of a Coronal Flux Rope}. \emph{\apj}
  \textbf{697}~(2), 1529--1542 (2009), \doi{10.1088/0004-637X/697/2/1529}.

\bibitem{Aulanier2010}
G.~{Aulanier}, T.~{T{\"o}r{\"o}k}, P.~{D{\'e}moulin}, E.~E. {DeLuca},
  {Formation of Torus-Unstable Flux Ropes and Electric Currents in Erupting
  Sigmoids}. \emph{\apj} \textbf{708}~(1), 314--333 (2010),
  \doi{10.1088/0004-637X/708/1/314}.

\bibitem{Antiochos2013}
S.~K. {Antiochos}, {Helicity Condensation as the Origin of Coronal and Solar
  Wind Structure}. \emph{\apj} \textbf{772}~(1), 72 (2013),
  \doi{10.1088/0004-637X/772/1/72}.

\bibitem{Liu2022}
Q.~{Liu}, C.~{Xia}, {Formation of Quiescent Prominence Magnetic Fields by
  Supergranulations}. \emph{\apjl} \textbf{934}~(1), L9 (2022),
  \doi{10.3847/2041-8213/ac80c6}.

\bibitem{Patsourakos2020}
S.~{Patsourakos}, \emph{et~al.}, {Decoding the Pre-Eruptive Magnetic Field
  Configurations of Coronal Mass Ejections}. \emph{\ssr} \textbf{216}~(8), 131
  (2020), \doi{10.1007/s11214-020-00757-9}.

\bibitem{Somov1998}
B.~V. {Somov}, T.~{Kosugi}, T.~{Sakao}, {Collisionless Three-dimensional
  Reconnection In Impulsive Solar Flares}. \emph{\apj} \textbf{497}~(2),
  943--956 (1998), \doi{10.1086/305492}.

\bibitem{Aulanier2021}
G.~{Aulanier}, {The return of the jet}. \emph{Nature Astronomy} \textbf{5},
  1096--1097 (2021), \doi{10.1038/s41550-021-01416-x}.

\bibitem{Antiochos1998}
S.~K. {Antiochos}, {The Magnetic Topology of Solar Eruptions}. \emph{\apjl}
  \textbf{502}~(2), L181--L184 (1998), \doi{10.1086/311507}.

\bibitem{Aulanier2019}
G.~{Aulanier}, J.~{Dud{\'\i}k}, {Drifting of the line-tied footpoints of CME
  flux-ropes}. \emph{\aap} \textbf{621}, A72 (2019),
  \doi{10.1051/0004-6361/201834221}.

\bibitem{Wang2002}
H.~{Wang}, \emph{et~al.}, {Rapid Changes of Magnetic Fields Associated with Six
  X-Class Flares}. \emph{\apj} \textbf{576}~(1), 497--504 (2002),
  \doi{10.1086/341735}.

\bibitem{Zharkova2005}
V.~V. {Zharkova}, S.~I. {Zharkov}, S.~S. {Ipson}, A.~K. {Benkhalil}, {Toward
  magnetic field dissipation during the 23 July 2002 solar flare measured with
  Solar and Heliospheric Observatory/Michelson Doppler Imager (SOHO/MDI) and
  Reuven Ramaty High Energy Solar Spectroscopic Imager (RHESSI)}. \emph{Journal
  of Geophysical Research (Space Physics)} \textbf{110}~(A8), A08104 (2005),
  \doi{10.1029/2004JA010934}.

\bibitem{Petrie2012}
G.~J.~D. {Petrie}, {The Abrupt Changes in the Photospheric Magnetic and Lorentz
  Force Vectors during Six Major Neutral-line Flares}. \emph{\apj}
  \textbf{759}~(1), 50 (2012), \doi{10.1088/0004-637X/759/1/50}.

\bibitem{Sudol2005}
J.~J. {Sudol}, J.~W. {Harvey}, {Longitudinal Magnetic Field Changes
  Accompanying Solar Flares}. \emph{\apj} \textbf{635}~(1), 647--658 (2005),
  \doi{10.1086/497361}.

\bibitem{Wang2012}
H.~{Wang}, N.~{Deng}, C.~{Liu}, {Rapid Transition of Uncombed Penumbrae to
  Faculae during Large Flares}. \emph{\apj} \textbf{748}~(2), 76 (2012),
  \doi{10.1088/0004-637X/748/2/76}.

\bibitem{Sun2017}
X.~{Sun}, J.~T. {Hoeksema}, Y.~{Liu}, M.~{Kazachenko}, R.~{Chen},
  {Investigating the Magnetic Imprints of Major Solar Eruptions with SDO/HMI
  High-cadence Vector Magnetograms}. \emph{\apj} \textbf{839}~(1), 67 (2017),
  \doi{10.3847/1538-4357/aa69c1}.

\bibitem{Bi2016}
Y.~{Bi}, \emph{et~al.}, {Observation of a reversal of rotation in a sunspot
  during a solar flare}. \emph{Nature Communications} \textbf{7}, 13798 (2016),
  \doi{10.1038/ncomms13798}.

\bibitem{Liu2016}
C.~{Liu}, \emph{et~al.}, {Flare differentially rotates sunspot on Sun's
  surface}. \emph{Nature Communications} \textbf{7}, 13104 (2016),
  \doi{10.1038/ncomms13104}.

\bibitem{Hudson2008}
H.~S. {Hudson}, G.~H. {Fisher}, B.~T. {Welsch}, {Flare Energy and Magnetic
  Field Variations}, in \emph{Subsurface and Atmospheric Influences on Solar
  Activity}, R.~{Howe}, R.~W. {Komm}, K.~S. {Balasubramaniam}, G.~J.~D.
  {Petrie}, Eds., vol. 383 of \emph{Astronomical Society of the Pacific
  Conference Series} (2008), p. 221.

\bibitem{Lau1990}
Y.-T. {Lau}, J.~M. {Finn}, {Three-dimensional Kinematic Reconnection in the
  Presence of Field Nulls and Closed Field Lines}. \emph{\apj} \textbf{350},
  672 (1990), \doi{10.1086/168419}.

\bibitem{Masson2009}
S.~{Masson}, E.~{Pariat}, G.~{Aulanier}, C.~J. {Schrijver}, {The Nature of
  Flare Ribbons in Coronal Null-Point Topology}. \emph{\apj} \textbf{700}~(1),
  559--578 (2009), \doi{10.1088/0004-637X/700/1/559}.

\bibitem{Yang2020}
S.~{Yang}, \emph{et~al.}, {Imaging and Spectral Study on the Null Point of a
  Fan-spine Structure During a Solar Flare}. \emph{\apj} \textbf{898}~(2), 101
  (2020), \doi{10.3847/1538-4357/ab9ac7}.

\bibitem{Cheng2011}
X.~{Cheng}, J.~{Zhang}, Y.~{Liu}, M.~D. {Ding}, {Observing Flux Rope Formation
  During the Impulsive Phase of a Solar Eruption}. \emph{\apjl}
  \textbf{732}~(2), L25 (2011), \doi{10.1088/2041-8205/732/2/L25}.

\bibitem{Zhang2012}
J.~{Zhang}, X.~{Cheng}, M.-D. {Ding}, {Observation of an evolving magnetic flux
  rope before and during a solar eruption}. \emph{Nature Communications}
  \textbf{3}, 747 (2012), \doi{10.1038/ncomms1753}.

\bibitem{Xing2024a}
C.~{Xing}, G.~{Aulanier}, B.~{Schmieder}, X.~{Cheng}, M.~{Ding}, {Identifying
  footpoints of pre-eruptive and coronal mass ejection flux ropes with sunspot
  scars}. \emph{\aap} \textbf{682}, A3 (2024),
  \doi{10.1051/0004-6361/202347053}.

\bibitem{Solanki2003}
S.~K. {Solanki}, {Sunspots: An overview}. \emph{\aapr} \textbf{11}~(2-3),
  153--286 (2003), \doi{10.1007/s00159-003-0018-4}.

\bibitem{Toriumi2015}
S.~{Toriumi}, Y.~{Katsukawa}, M.~C.~M. {Cheung}, {Light Bridge in a Developing
  Active Region. I. Observation of Light Bridge and its Dynamic Activity
  Phenomena}. \emph{\apj} \textbf{811}~(2), 137 (2015),
  \doi{10.1088/0004-637X/811/2/137}.

\bibitem{Janvier2014}
M.~{Janvier}, \emph{et~al.}, {Electric Currents in Flare Ribbons: Observations
  and Three-dimensional Standard Model}. \emph{\apj} \textbf{788}~(1), 60
  (2014), \doi{10.1088/0004-637X/788/1/60}.

\bibitem{Titov1999}
V.~S. {Titov}, P.~{D{\'e}moulin}, {Basic topology of twisted magnetic
  configurations in solar flares}. \emph{\aap} \textbf{351}, 707--720 (1999).

\bibitem{Fisher2012}
G.~H. {Fisher}, D.~J. {Bercik}, B.~T. {Welsch}, H.~S. {Hudson}, {Global Forces
  in Eruptive Solar Flares: The Lorentz Force Acting on the Solar Atmosphere
  and the Solar Interior}. \emph{\solphys} \textbf{277}~(1), 59--76 (2012),
  \doi{10.1007/s11207-011-9907-2}.

\bibitem{Munoz2018}
P.~A. {Mu{\~n}oz}, J.~{B{\"u}chner}, {Kinetic turbulence in fast
  three-dimensional collisionless guide-field magnetic reconnection}.
  \emph{\pre} \textbf{98}~(4), 043205 (2018), \doi{10.1103/PhysRevE.98.043205}.

\bibitem{Khabarova2020}
O.~{Khabarova}, V.~{Zharkova}, Q.~{Xia}, O.~E. {Malandraki}, {Counterstreaming
  Strahls and Heat Flux Dropouts as Possible Signatures of Local Particle
  Acceleration in the Solar Wind}. \emph{\apjl} \textbf{894}~(1), L12 (2020),
  \doi{10.3847/2041-8213/ab8cb8}.

\bibitem{Zharkova2021}
V.~{Zharkova}, Q.~{Xia}, {Plasma turbulence generated in 3D current sheets with
  single and multiple X-nullpoints}. \emph{Frontiers in Astronomy and Space
  Sciences} \textbf{8}, 178 (2021), \doi{10.3389/fspas.2021.665998}.

\bibitem{Barczynski2019}
K.~{Barczynski}, G.~{Aulanier}, S.~{Masson}, M.~S. {Wheatland}, {Flare
  Reconnection-driven Magnetic Field and Lorentz Force Variations at the
  Sun{\textquoteright}s Surface}. \emph{\apj} \textbf{877}~(2), 67 (2019),
  \doi{10.3847/1538-4357/ab1b3d}.

\bibitem{Aulanier2016}
G.~{Aulanier}, {Solar physics: When the tail wags the dog}. \emph{Nature
  Physics} \textbf{12}~(11), 998--999 (2016), \doi{10.1038/nphys3938}.

\bibitem{Castellanos2018}
J.~S. {Castellanos Dur{\'a}n}, L.~{Kleint}, B.~{Calvo-Mozo}, {A Statistical
  Study of Photospheric Magnetic Field Changes During 75 Solar Flares}.
  \emph{\apj} \textbf{852}~(1), 25 (2018), \doi{10.3847/1538-4357/aa9d37}.

\bibitem{Pariat2009}
E.~{Pariat}, S.~K. {Antiochos}, C.~R. {DeVore}, {A Model for Solar Polar Jets}.
  \emph{\apj} \textbf{691}~(1), 61--74 (2009),
  \doi{10.1088/0004-637X/691/1/61}.

\bibitem{Torok2009}
T.~{T{\"o}r{\"o}k}, G.~{Aulanier}, B.~{Schmieder}, K.~K. {Reeves}, L.~{Golub},
  {Fan-Spine Topology Formation Through Two-Step Reconnection Driven by Twisted
  Flux Emergence}. \emph{\apj} \textbf{704}~(1), 485--495 (2009),
  \doi{10.1088/0004-637X/704/1/485}.

\bibitem{Wyper2016}
P.~F. {Wyper}, C.~R. {DeVore}, J.~T. {Karpen}, B.~J. {Lynch},
  {Three-Dimensional Simulations of Tearing and Intermittency in Coronal Jets}.
  \emph{\apj} \textbf{827}~(1), 4 (2016), \doi{10.3847/0004-637X/827/1/4}.

\bibitem{Yang2009}
S.~{Yang}, J.~{B{\"u}chner}, H.~{Zhang}, {Magnetic Helicity Exchange Between
  Neighboring Active Regions}. \emph{\apjl} \textbf{695}~(1), L25--L30 (2009),
  \doi{10.1088/0004-637X/695/1/L25}.

\bibitem{Drake2000}
J.~J. {Drake}, G.~{Peres}, S.~{Orlando}, J.~M. {Laming}, A.~{Maggio}, {On
  Stellar Coronae and Solar Active Regions}. \emph{\apj} \textbf{545}~(2),
  1074--1083 (2000), \doi{10.1086/317820}.

\bibitem{Namekata2022}
K.~{Namekata}, \emph{et~al.}, {Probable detection of an eruptive filament from
  a superflare on a solar-type star}. \emph{Nature Astronomy} \textbf{6},
  241--248 (2021), \doi{10.1038/s41550-021-01532-8}.

\bibitem{Lemen2012}
J.~R. {Lemen}, \emph{et~al.}, {The Atmospheric Imaging Assembly (AIA) on the
  Solar Dynamics Observatory (SDO)}. \emph{\solphys} \textbf{275}~(1-2), 17--40
  (2012), \doi{10.1007/s11207-011-9776-8}.

\bibitem{Bobra2014}
M.~G. {Bobra}, \emph{et~al.}, {The Helioseismic and Magnetic Imager (HMI)
  Vector Magnetic Field Pipeline: SHARPs - Space-Weather HMI Active Region
  Patches}. \emph{\solphys} \textbf{289}~(9), 3549--3578 (2014),
  \doi{10.1007/s11207-014-0529-3}.

\bibitem{Scherrer2012}
P.~H. {Scherrer}, \emph{et~al.}, {The Helioseismic and Magnetic Imager (HMI)
  Investigation for the Solar Dynamics Observatory (SDO)}. \emph{\solphys}
  \textbf{275}~(1-2), 207--227 (2012), \doi{10.1007/s11207-011-9834-2}.

\bibitem{Pesnell2012}
W.~D. {Pesnell}, B.~J. {Thompson}, P.~C. {Chamberlin}, {The Solar Dynamics
  Observatory (SDO)}. \emph{\solphys} \textbf{275}~(1-2), 3--15 (2012),
  \doi{10.1007/s11207-011-9841-3}.

\bibitem{Brueckner1995}
G.~E. {Brueckner}, \emph{et~al.}, {The Large Angle Spectroscopic Coronagraph
  (LASCO)}. \emph{\solphys} \textbf{162}~(1-2), 357--402 (1995),
  \doi{10.1007/BF00733434}.

\bibitem{Domingo1995}
V.~{Domingo}, B.~{Fleck}, A.~I. {Poland}, {The SOHO Mission: an Overview}.
  \emph{\solphys} \textbf{162}~(1-2), 1--37 (1995), \doi{10.1007/BF00733425}.

\bibitem{Xia2018}
C.~{Xia}, J.~{Teunissen}, I.~{El Mellah}, E.~{Chan{\'e}}, R.~{Keppens},
  {MPI-AMRVAC 2.0 for Solar and Astrophysical Applications}. \emph{\apjs}
  \textbf{234}~(2), 30 (2018), \doi{10.3847/1538-4365/aaa6c8}.

\bibitem{Dedner2002}
A.~{Dedner}, \emph{et~al.}, {Hyperbolic Divergence Cleaning for the MHD
  Equations}. \emph{Journal of Computational Physics} \textbf{175}~(2),
  645--673 (2002), \doi{10.1006/jcph.2001.6961}.

\bibitem{Jiang1996}
G.-S. {Jiang}, C.-W. {Shu}, {Efficient Implementation of Weighted ENO Schemes}.
  \emph{Journal of Computational Physics} \textbf{126}~(1), 202--228 (1996),
  \doi{10.1006/jcph.1996.0130}.

\bibitem{Shu2009}
C.-W. {Shu}, {High Order Weighted Essentially Nonoscillatory Schemes for
  Convection Dominated Problems}. \emph{SIAM Review} \textbf{51}~(1), 82--126
  (2009), \doi{10.1137/070679065}.

\bibitem{Tanaka1994}
T.~{Tanaka}, {Finite Volume TVD Scheme on an Unstructured Grid System for
  Three-Dimensional MHD Simulation of Inhomogeneous Systems Including Strong
  Background Potential Fields}. \emph{Journal of Computational Physics}
  \textbf{111}~(2), 381--389 (1994), \doi{10.1006/jcph.1994.1071}.

\bibitem{Xing2024b}
C.~{Xing}, G.~{Aulanier}, X.~{Cheng}, C.~{Xia}, M.~{Ding}, {Unveiling the
  Initiation Route of Coronal Mass Ejections through Their Slow Rise Phase}.
  \emph{\apj} \textbf{966}~(1), 70 (2024), \doi{10.3847/1538-4357/ad2ea9}.

\bibitem{Jiang2021}
C.~{Jiang}, \emph{et~al.}, {A fundamental mechanism of solar eruption
  initiation}. \emph{Nature Astronomy} \textbf{5}, 1126--1138 (2021),
  \doi{10.1038/s41550-021-01414-z}.

\bibitem{Priest1995}
E.~R. {Priest}, P.~{D{\'e}moulin}, {Three-dimensional magnetic reconnection
  without null points. 1. Basic theory of magnetic flipping}. \emph{\jgr}
  \textbf{100}~(A12), 23443--23464 (1995), \doi{10.1029/95JA02740}.

\bibitem{Titov2002}
V.~S. {Titov}, G.~{Hornig}, P.~{D{\'e}moulin}, {Theory of magnetic connectivity
  in the solar corona}. \emph{Journal of Geophysical Research (Space Physics)}
  \textbf{107}~(A8), 1164 (2002), \doi{10.1029/2001JA000278}.

\bibitem{Zhang2022}
P.~{Zhang}, J.~{Chen}, R.~{Liu}, C.~{Wang}, {FastQSL: A Fast Computation Method
  for Quasi-separatrix Layers}. \emph{\apj} \textbf{937}~(1), 26 (2022),
  \doi{10.3847/1538-4357/ac8d61}.

\end{thebibliography}
\bibliographystyle{sciencemag}


\section*{Acknowledgments}
We thank Can Wang and Bo Li for valuable discussions.
The numerical calculation was done on the computing facilities in the High Performance Computing Center of Nanjing University.
\paragraph*{Funding:}
The work of C. X. and X. C. was supported by the NSFC under grants 12403066 and 12525305, the Jiangsu NSF under grant BK20241187, the Fundamental Research Funds for the Central Universities under grants 2024300348 and 2025300318, and the Jiangsu Funding Program for Excellent Postdoctoral Talent.
The work of G. A. was supported by the Partenariat Hubert Curien (PHC) Cai Yuanpei for scientific cooperation between France and China, as well by the Appel \'a Proposition de Recherche of CNES/SHM and by the Action Th\'ematique Soleil-Terre (ATST) of CNRS/INSU PN Astro, also funded by CNES, CEA, and ONERA.
\paragraph*{Author contributions:}
Conceptualization: CX, XC, GA, MDD \\
Methodology: CX, XC \\
Software: CX \\
Validation: CX \\
Formal analysis: CX \\
Investigation: CX \\
Resources: CX, XC \\
Data curation: CX \\
Writing - original draft: CX, XC \\
Writing - review \& editing: CX, XC, GA, MDD \\
Visualization: CX \\
Supervision: XC, MDD \\
Project administration: CX, XC, MDD \\
Funding acquisition: CX, XC, GA
\paragraph*{Competing interests:}
The authors declare they have no competing interest.
\paragraph*{Data and materials availability:}
The observation data by SDO/AIA and SDO/HMI are available at the website \url{http://jsoc.stanford.edu/ajax/exportdata.html}. One can obtain the data by following the instructions on the website.
The simulation data presented in this work is archived at the website \url{https://sdc.nju.edu.cn/d/eb2177272eec4f08ac7d/} managed by the Solar Data Center of Nanjing University.
This study did not generate new materials.
MPI-AMRVAC 2.2, the code used to perform the simulation in this work, is an open-source code. It is available from the website \url{https://github.com/amrvac/amrvac/releases/tag/v2.2}. The code used to calculate the squashing degree is available on the website \url{https://github.com/el2718/FastQSL}.

\begin{figure}
\centering
\includegraphics[width=0.88\hsize]{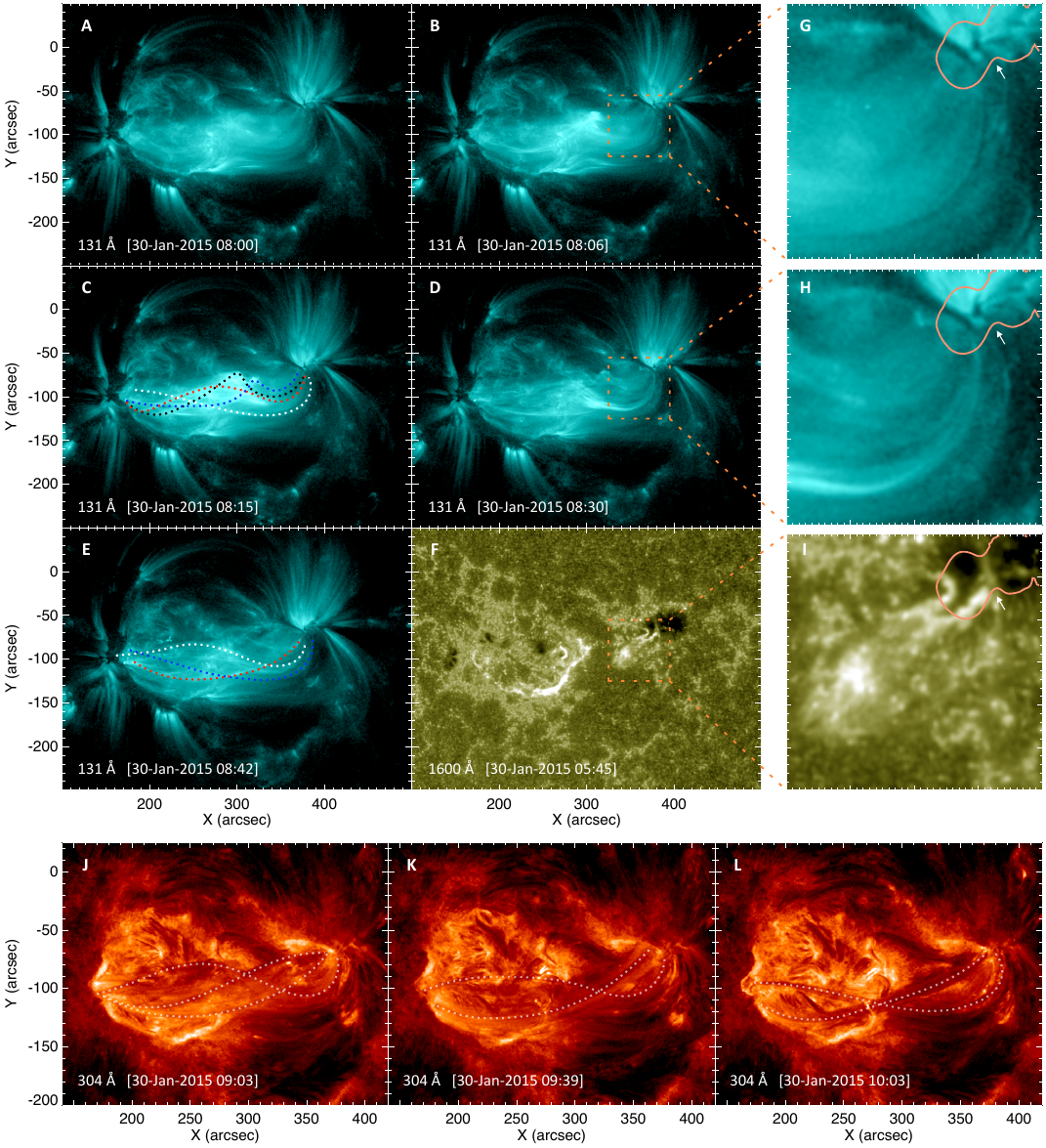}
\caption{\textbf{Temporal evolution of the twisted overlying loops.} (\textbf{A})-(\textbf{E}) Temporal evolution of the twisted overlying loops in 131 \AA\ after the eruption. The colored dashed curves in panels C and E sketch the twisted loops. (\textbf{F}) The 1600 \AA\ image of ARs 12268/70 during the eruption. (\textbf{G}) Zoom-in image of panel B showing the western footpoint of the twisted overlying loops. The FOV is marked by the dashed box in panel B. The orange curve is the contour of $B_z=1000$ G, and the white arrow points to the location of sunspot scar. (\textbf{H}) Similar to panel G, but showing the zoom-in image of panel D. (\textbf{I}) Similar to panel G, but showing the zoom-in image of panel F. (\textbf{J})-(\textbf{L}) Similar to panels A-E but showing the evolution of twisted overlying loops in 304 \AA. The dashed curves sketch the twisted loops.}
\label{fig2}
\end{figure}

\begin{figure}
\centering
\includegraphics[width=\hsize]{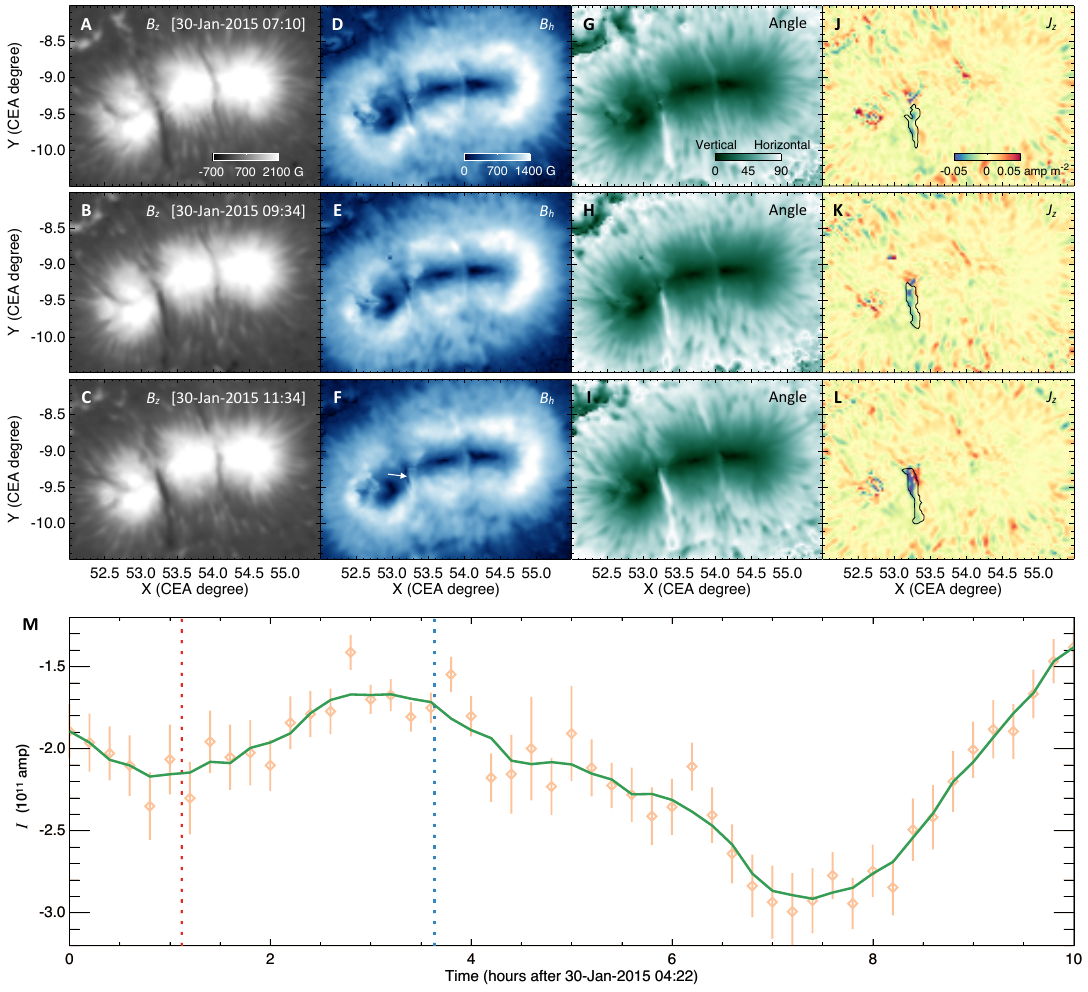}
\caption{\textbf{Temporal evolution of the sunspot scar after the eruption.} (\textbf{A})-(\textbf{C}) The distributions of $B_z$ in AR 12270 at three moments. (\textbf{D})-(\textbf{F}) Similar to panels A-C but showing the distributions of $B_h$. The white arrow points to the enhancement of $B_h$ in the sunspot scar. (\textbf{G})-(\textbf{I}) Similar to panels A-C but showing the distributions of the inclination angle of magnetic field to the vertical direction. (\textbf{J})-(\textbf{L}) Similar to panels A-C but showing the distributions of $J_z$. The black contours outline the negative-current region of the sunspot scar. (\textbf{M}) The evolution of the integral negative current ($I$) of the sunspot scar. The diamonds display the measured data with the error bars showing their uncertainties ($\sigma_I$). The curve shows a smooth profile of measurements. The red dashed line marks the eruption onset, and the blue dashed line marks the moment when the twisted overlying loops first appeared in 131 \AA\ after the eruption.}
\label{fig3}
\end{figure}

\begin{figure}
\centering
\includegraphics[width=\hsize]{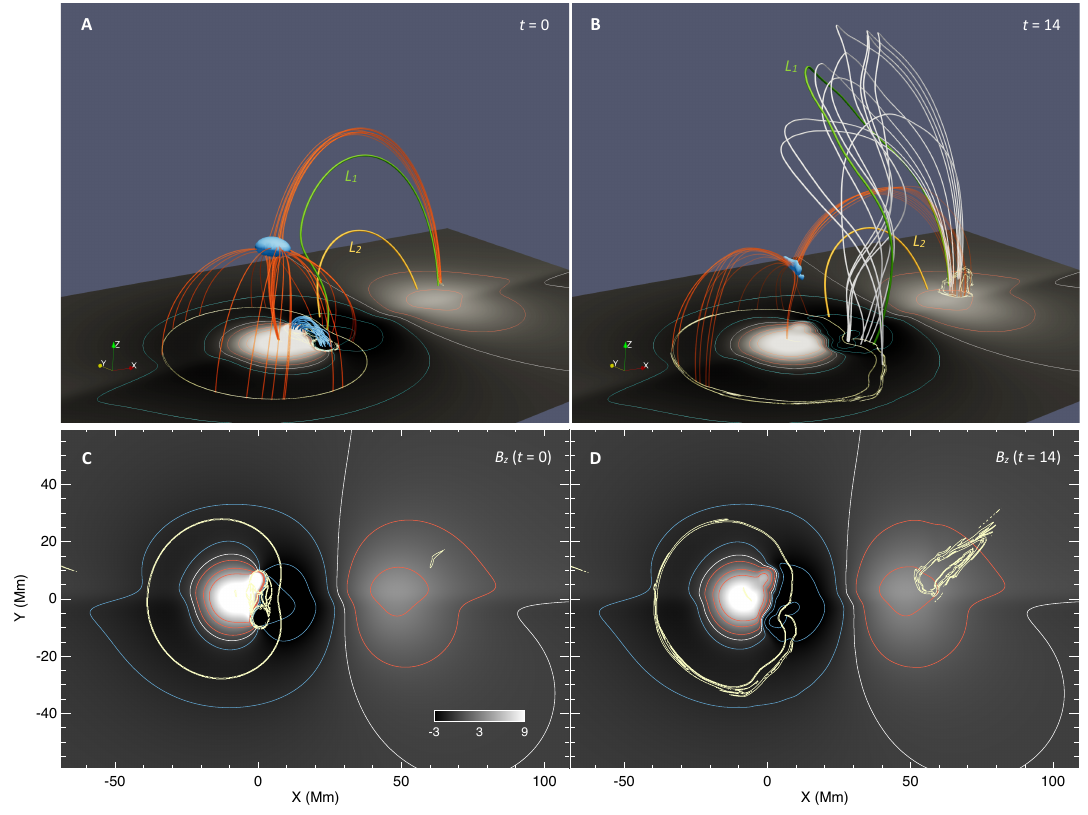} 
\caption{\textbf{MHD simulation of the failed eruption.} (\textbf{A}) The initial magnetic configuration of the simulation. The orange field lines represent those near the fan separatrix surface and inner/outer spine. The blue field lines represent the modified TD flux rope below the fan separatrix surface. The blue isosuface marks the low field-strength region around the null point. The green (yellow) field line represents $L_1$ ($L_2$). (\textbf{B}) The magnetic configuration after the failed eruption. The orange field lines show the remaining fan-spine structure, and the blue isosuface outlines the low field-strength region. The white field lines represent the twisted overlying loops connecting the remote positive polarity and the negative polarity. The green (yellow) field line represents $L_1$ ($L_2$). (\textbf{C}) The bottom surface in panel A, showing the distributions of $B_z$ on the observation layer (materials and methods S4). The red/blue curves are the contours of $B_z$ in the positive/negative polarities. The white curves mark the PILs. The yellow curves mark the boundaries of the footprints of quasi-separatrix layers (QSLs; \cite{Priest1995}), where QSLs are defined as regions with a squashing degree \cite{Titov2002,Zhang2022} greater than 1000. (\textbf{D}) Similar to panel C but showing the bottom surface in panel B.}
\label{fig4}
\end{figure}

\begin{figure}
\centering
\includegraphics[width=\hsize]{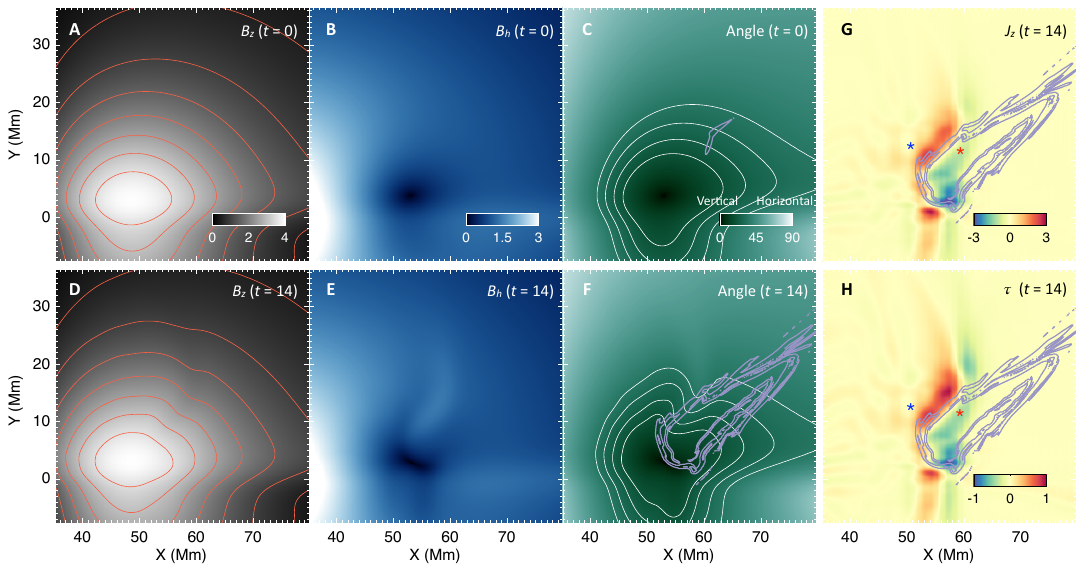}
\caption{\textbf{Modeled sunspot scar in the remote positive polarity.} (\textbf{A})-(\textbf{C}) The distributions of $B_z$, $B_h$, and inclination angle in the remote positive polarity on the observation layer at $t=0$. The red curves in panel A are the contours of $B_z$, and the white curves in panel C are the contours of inclination angle. (\textbf{D})-(\textbf{F}) Similar to panels A-C but at $t=14$. (\textbf{G})-(\textbf{H}) The distributions of $J_z$ and $\tau$ in the remote positive polarity on the observation layer at $t=14$. The red (blue) asterisk marks the footpoint of $L_1$ ($L_2$) on this layer. The purple curves in panels C and F-H represent the boundaries of the footprints of QSLs.}
\label{fig5}
\end{figure}


\subsection*{Supplementary materials}
Supplementary Text\\
Figs. S1 to S4\\


\newpage


\renewcommand{\thefigure}{S\arabic{figure}}
\renewcommand{\thetable}{S\arabic{table}}
\renewcommand{\theequation}{S\arabic{equation}}
\renewcommand{\thepage}{S\arabic{page}}
\setcounter{figure}{0}
\setcounter{table}{0}
\setcounter{equation}{0}
\setcounter{page}{1} 


\begin{center}
\section*{Supplementary Materials for\\ \scititle}

Chen~Xing$^{\ast}$,
Xin~Cheng$^{\ast}$,
Guillaume~Aulanier,
Mingde~Ding\\ 
\small$^\ast$Corresponding author. Email: chenxing@nju.edu.cn\\
\small$^\ast$Corresponding author. Email: xincheng@nju.edu.cn
\end{center}

\subsubsection*{This PDF file includes:}
Supplementary Text\\
Figures S1 to S4\\

\newpage


\subsection*{Supplementary Text}
\subsubsection*{Relationship between Evolutions of Scar and Untwisting Coronal Loops}
One can find that the period of the untwisting of twisted coronal loops in 131 \AA\ and 304 \AA\ images (08:00-10:12 UT; Fig. \ref{fig2}) is shorter than that of the sunspot scar growth (07:34-11:34 UT; Fig. \ref{fig3}M). Nevertheless, it does not substantially affect our result that the untwisting of coronal fields leads to the growth of sunspot scar, considering the following reasons. First, the untwisting process likely occurred over a longer period covering 08:00-10:12 UT, while it was only visible in AIA images during 08:00-10:12 UT when the temperature of twisted coronal loops was within the instrument's observation temperature range. In fact, the temperature of twisted coronal loops does vary, as they initially appeared in 131 \AA\ and were subsequently observable only in 304 \AA\ (Fig. \ref{fig2}). Second, the end of the sunspot scar growth should be later than that of the untwisting phenomenon in the corona, because it takes time for the latter to cause the former through the propagation of Alfvén waves whose speed is especially low in the lower atmosphere. Third, when Alfvén waves (launched by the untwisting of coronal loops) reach the positive-polarity footpoint of loops, a part of waves is refracted into the photosphere, causing a change of magnetic fields there and forming the sunspot scar. However, the other part of waves is reflected back into the corona and could propagate back to and grow the sunspot scar after reflected again at the other footpoint of loops. Such a possible process could occur several times as the overlying loops gradual relax toward a mechanical equilibrium, which may also explain why the sunspot scar continues to grow after the end of the untwisting of coronal loops.


\begin{figure}
\centering
\includegraphics[width=\hsize]{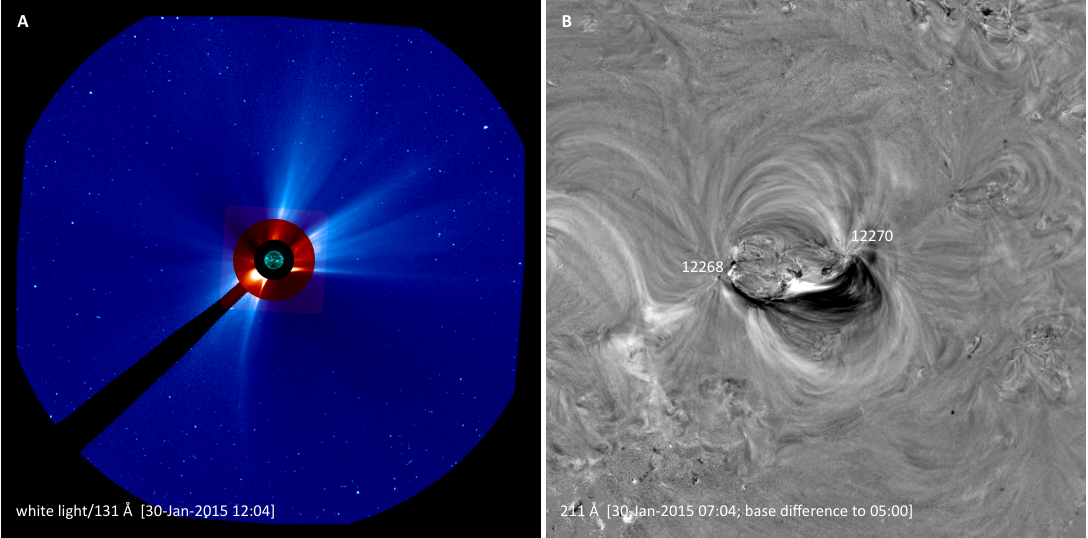}
\caption{\textbf{Evidence for the failure of the eruption.} (\textbf{A}) Composite image of the white-light images of LASCO C2 and C3 and the AIA 131 \AA\ image around 12:04 UT on 30 January 2015. (\textbf{B}) The base-difference image of the region around ARs 12268/70 in 211 \AA\ at 07:04 UT on 30 January 2015, with the base image taken at 05:00 UT.}
\label{figA1}
\end{figure}

\begin{figure}
\centering
\includegraphics[width=\hsize]{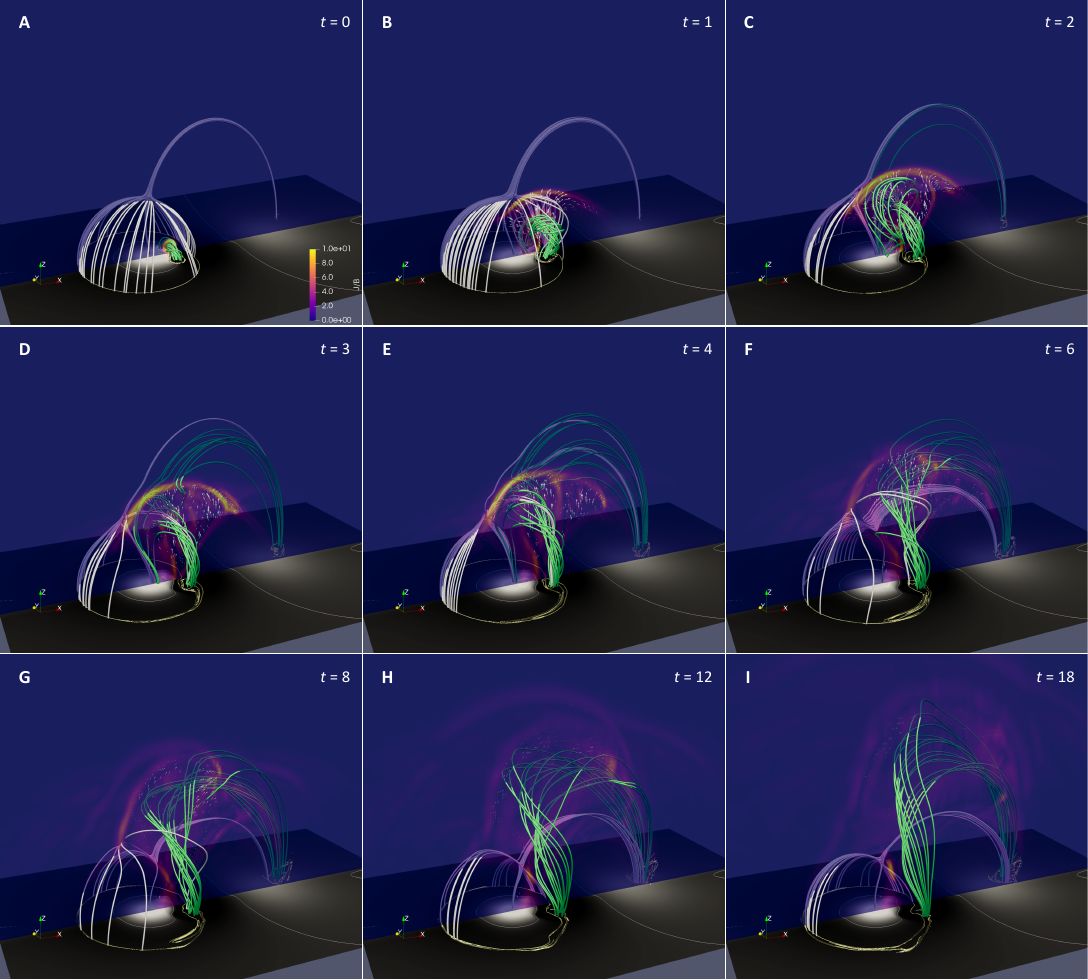}
\caption{\textbf{Evolution of the modeled failed eruption.} The green field lines show the pre-eruptive/erupting structure. The white field lines, traced from the low field-strength region, represent the (remaining) fan-spine structure. The vertical surface, $y=0$, shows the distribution of $J/B$. The white arrows represent the flow $v_x\boldsymbol{e}_x+v_z\boldsymbol{e}_z$ at this surface, with the length scales of arrows in each panel being the same. The bottom surface shows the distribution of $B_z$ on the observation layer, where the yellow contours represent the boundaries of the footprints of QSLs and the white curves represent the PILs.}
\label{figA2}
\end{figure}

\begin{figure}
\centering
\includegraphics[width=\hsize]{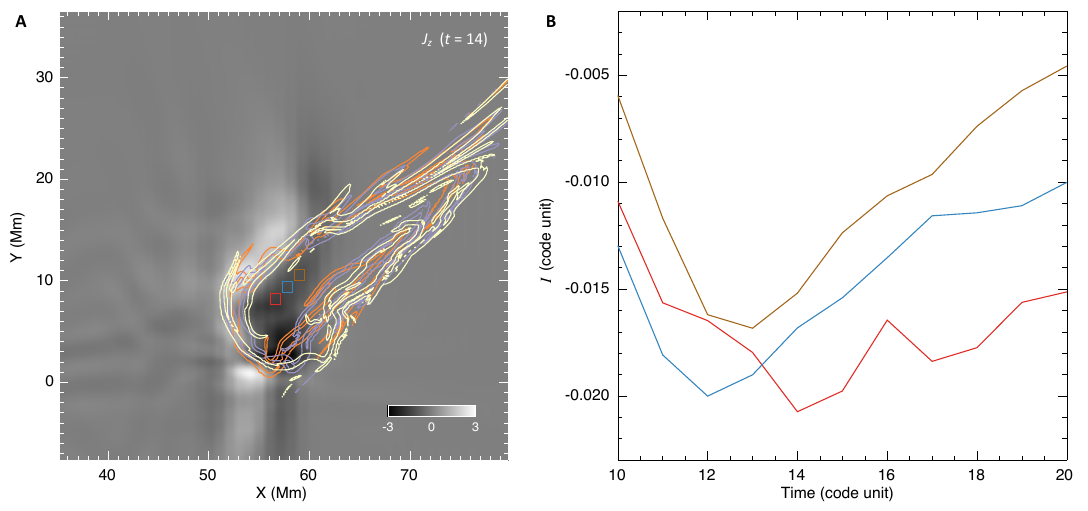}
\caption{\textbf{Evolution of the current of the modeled sunspot scar.} (\textbf{A}) The distribution of $J_z$ in the remote positive polarity on the observation layer at $t=14$. The orange, purple, and yellow curves represent the boundaries of the footprints of QSLs at $t=10$, $t=14$, and $t=20$, respectively. (\textbf{B}) The red, blue, and brown curves represent the evolutions of the currents ($I$) integrated in the red, blue, and brown boxes in panel A, respectively, the latter of which are always located in the negative-current region of sunspot scar during $10\le t\le20$.}
\label{figA3}
\end{figure}

\begin{figure}
\centering
\includegraphics[width=\hsize]{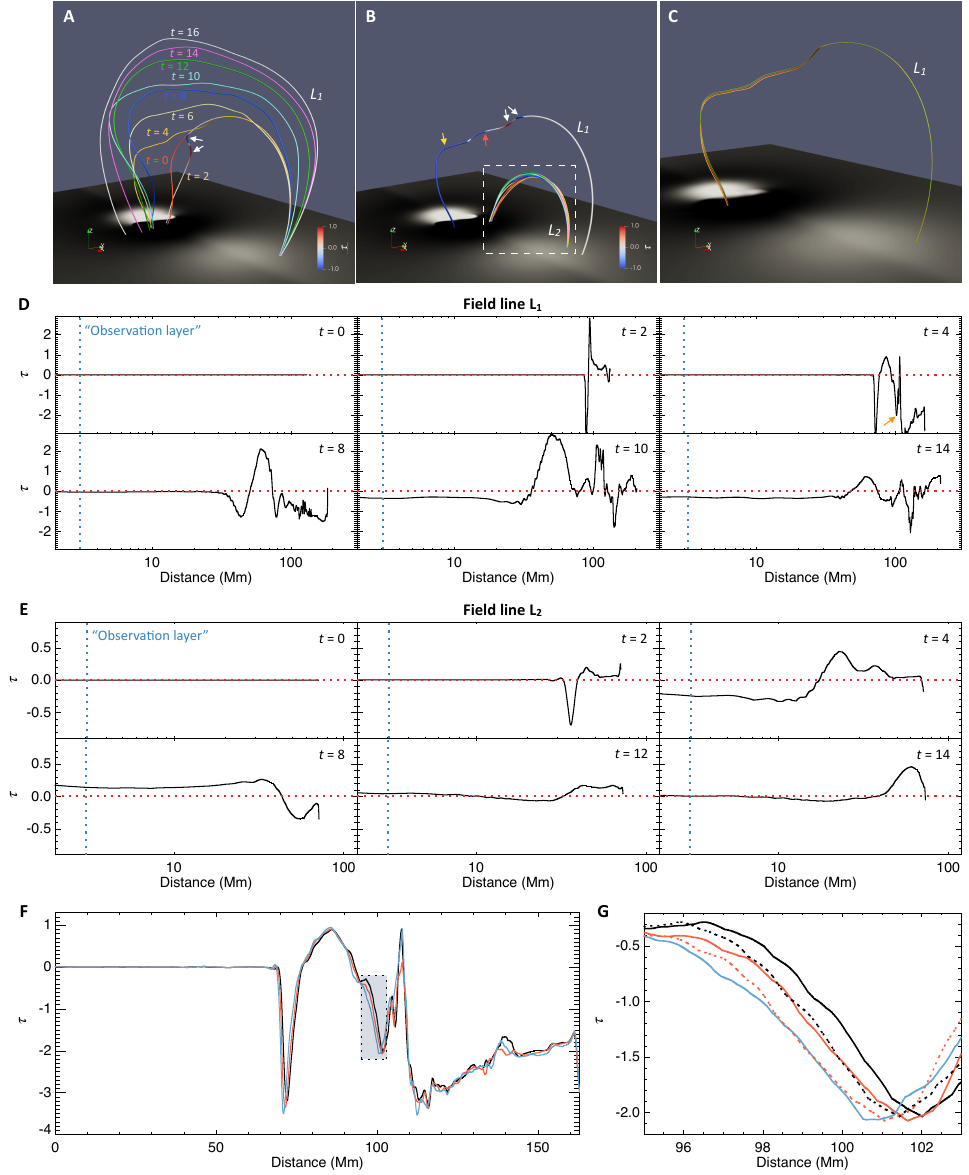}
\label{figA4}
\end{figure}

\begin{figure}
\centering
\caption{\textbf{Evolutions of the field lines anchored around the sunspot scar.} (\textbf{A}) Evolution of $L_1$ during $0\le t\le16$. The color of the field line at $t=2$ shows the distribution of $\tau$ along $L_1$, and the white arrows point to a pair of disturbance-induced twist. The bottom surfaces exhibit the distributions of $B_z$ on the observation layer. (\textbf{B}) Similar to panel A but showing the evolution of $L_2$ (in the dashed box). In addition, the field line outside the dashed box represents $L_1$ at $t=4$, and its color shows the distribution of $\tau$ along it. The white arrows point to the disturbance-induced twist. The yellow arrow points to the reconnection-induced twist, with the red arrow marking its leading front. (\textbf{C}) Evolution of $L_1$ from $t=4$ (yellow) to $t=4.05$ (red) to $t=4.1$ (green). (\textbf{D}) Evolution of the distribution of $\tau$ along $L_1$. The distance refers to that along $L_1$ to its positive-polarity footpoint on the line-tied layer. The red dashed line marks $\tau=0$, and the blue dashed line marks the intersection of $L_1$ and the observation layer. The yellow arrow points to the reconnection-induced twist. (\textbf{E}) Similar to panel D but showing the evolution of $\tau$ along $L_2$. (\textbf{F}) Evolution of the distribution of $\tau$ along $L_1$ from $t=4$ (black) to $t=4.05$ (red) to $t=4.1$ (blue). (\textbf{G}) The zoom-in image of the gray region in panel F. The solid curves have the same meanings as those in panel F. The black (red) dashed curve represents the shifted distribution of $\tau$ at $t=4.05$ ($t=4.1$).}
\end{figure}

\end{document}